\documentclass{bioRxiv}

\usepackage{url}
\usepackage{graphicx}
\usepackage{cmbright}
\usepackage{fancyvrb}
\usepackage{indentfirst}
\usepackage[font={small}]{caption}

\begin{document}

\leadauthor{Matelsky}

\title{A large language model-assisted education tool to provide feedback on open-ended responses}
\shorttitle{Harnessing large language models for education}

\author[1,2]{Jordan K. Matelsky \orcidlink{0000-0002-9470-760X}}
\author[3]{Felipe Parodi \orcidlink{0000-0001-8948-490X}}
\author[4]{Tony Liu \orcidlink{0000-0002-3707-3989}}
\author[1,5]{Richard D. Lange \orcidlink{0000-0002-1429-8333}}
\author[1,3,4,6]{Konrad P. Kording \orcidlink{0000-0001-8408-4499}}
\affil[1]{Department of Bioengineering, University of Pennsylvania}
\affil[2]{Research \& Exploratory Development Department, Johns Hopkins University Applied Physics Laboratory}
\affil[3]{Department of Neuroscience, University of Pennsylvania}
\affil[4]{Department of Computer Science, University of Pennsylvania}
\affil[5]{Department of Computer Science, Rochester Institute of Technology}
\affil[6]{CIFAR LMB Program}
\date{\today}

\maketitle

\begin{abstract}
Open-ended questions are a favored tool among instructors for assessing student understanding and encouraging critical exploration of course material. Providing feedback for such responses is a time-consuming task that can lead to overwhelmed instructors and decreased feedback quality. Many instructors resort to simpler question formats, like multiple-choice questions, which provide immediate feedback but at the expense of  personalized and insightful comments. Here, we present a tool that uses large language models (LLMs), guided by instructor-defined criteria, to automate responses to open-ended questions. Our tool delivers rapid personalized feedback, enabling students to quickly test their knowledge and identify areas for improvement. We provide open-source reference implementations both as a web application and as a Jupyter Notebook widget that can be used with instructional coding or math notebooks. With instructor guidance, LLMs hold promise to enhance student learning outcomes and elevate instructional methodologies.
\end{abstract}

\vspace{-.2em}
\begin{corrauthor}
matelsky@seas.upenn.edu
\end{corrauthor}

\section*{Introduction}

Open-ended questions --- questions that require students to produce multi-word, nontrivial responses --- are a popular assessment tool in educational environments because they offer students the chance to explore their understanding of learning material. Such questions provide valuable insight into students' grasp of complex concepts and their problem-solving approaches. However, grading open-ended questions can be time-consuming, subjective, and --- especially in the case of large class sizes --- prone to attentional errors. These factors create a critical bottleneck in precision education.

Large Language Models (LLMs) present an opportunity to automate and promote equity in learning assessments, providing rapid valuable feedback to students while reducing the burden on instructors. We developed a tool that automatically assesses students' responses to open-ended questions by evaluating their responses against a set of instructor-defined criteria.
To use our tool, the instructor poses a question along with optional grading criteria. Students respond to these questions, and their answers are relayed to a server. The responses are paired with the grading criteria (which are not revealed to the student), forming a payload for a large language model (LLM). The LLM then generates automated feedback, suggesting areas for improvement to the student.

Here, we describe the technical design of our tool, \textit{FreeText}, and showcase its utility in educational environments spanning topics and complexity. We further outline the implications of our work for teaching complex subjects, and the potential role of large language models in education (\textbf{Fig.~\ref{fig:sota})}. We share our source code and a public URL (see \textit{Supplemental Materials}), allowing educators to experiment with \textit{FreeText} firsthand.

\begin{figure}[!b]
    \includegraphics[width=\linewidth]{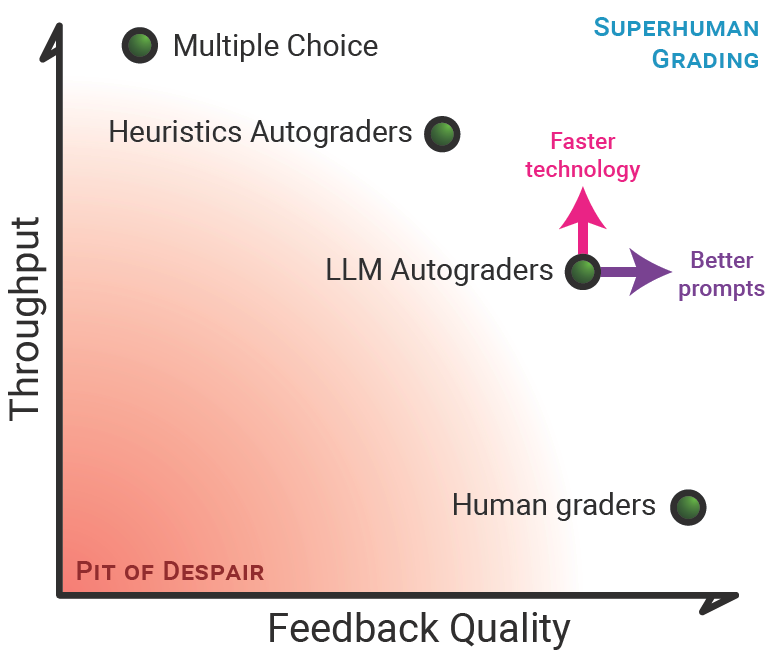}
    \caption{\textbf{
    Sketch comparing grading throughput and quality of feedback to students among various assessment methodologies} The $y$-axis represents throughput (i.e., rapidity of feedback generation and number of assignments evaluated per real-world unit-time or cost), and the $x$-axis represents feedback quality (a qualitative measure of personalization and detail of feedback given to students). LLMs have the potential to fill a niche among educational tools by striking a balance between quantity and quality, delivering high throughput with feedback quality comparable to human graders. Improvements in technology (faster GPU cards, better LLM architectures) will continue to push throughput upward, and improvements in prompt design (or other domain-specific adaptations) will improve the quality of LLM-generated feedback.}
    \label{fig:sota}
\end{figure}

\section*{Related Work}

Automated grading is a longstanding pursuit in the field of education technology. Early automated grading tools focused on `solvable' tasks like math or programming assignments, where grading generally relies on unit tests or direct output comparisons \citep{hollingsworth_automatic_1960, ureel_ii_automated_2019, orr_automatic_2021, messer_automated_2023}. These approaches often overlook less easily-quantified but nonetheless critical indicators of learning and understanding, such as design quality, code maintainability, or potential areas of student confusion. Modern tools, like AutoGrader, which provides real-time grading for programming exercises, remain narrowly focused on output correctness and do not sufficiently account for documentation or maintainability \citep{liu_automatic_2019}.

Assessing students' understanding from natural language responses, however, presents different challenges and has seen significant evolution. Early Automated Short Answer Grading (ASAG) models employed statistical or domain-specific neural network approaches \citep{heilman_ets_2013, riordan_investigating_2017, sung_pre-training_2019}. In recent years, LLMs have been shown to outperform domain-specific language models \citep{radford_language_2019, mizumoto_analytic_2019, brown_language_2020, chung_scaling_2022}. LLMs facilitate grading of open-ended assignment responses, without the need for task-specific fine-tuning \citep{cao_leveraging_2023, mizumoto_exploring_2023, yoon_short_2023}. However, \cite{kortemeyer_can_2023} revealed that while LLMs like GPT-4 could be useful for preliminary grading of introductory physics assignments, they fell short for natural-language responses required in comprehensive exam grading. Further, while LLMs like GitHub Copilot streamline the process of code generation and review, they can fall short on more nuanced programming tasks and open-ended evaluation~\citep{finnie2022robots}. Thus, in their current state, LLMs should be treated as a useful but fallible tool, with final assessments still in the hands of (human) instructors.

It is also important to consider how students perceive AI graders and how automated graders are deployed to educational settings \citep{burrows2015eras,saha2019joint,zhu2022automatic}. Many comment on the socio-technical dynamics of automated grading, including the potential for introduction of machine bias (e.g., \cite{hsu2021attitudes}). The use of NLP for short answer grading is not a trivial task and has been set as an evaluation challenge in its own right \citep{dzikovska2013semeval}.

To address the evolving needs of grading open-ended responses, our framework proposes four key enhancements. First, it is specifically designed for open-ended questions, which are not typically well-served by the rubric-based grading of most ed-tech tools. Second, our system leverages LLMs to deliver rapid, personalized feedback for student responses without explicitly attempting to produce a quantitative grade. Third, our framework introduces a feedback loop to continually improve instructor-provided prompts, question suggestions, and grading criteria. Lastly, our tool integrates with the Jupyter Notebook environment, extensively utilized in fields such as computer science, data science, and statistics.

\section*{Approach}

We have designed our tool for use in a variety of educational contexts, ranging from primary school education to graduate courses. \textit{FreeText} enables educators to integrate open-ended questions into their curriculum without incurring an instructor labor cost. This allows students to gain rapid, individualized, and sophisticated feedback, thereby creating a highly effective learning loop that can enhance the absorption of course materials. It guides students in refining their responses, enhancing their understanding and application of concepts in each iteration. This feedback is generated by a large language model (LLM), which circumvents the attentional errors often made by human graders, particularly when assessing a large volume of assignments. The LLM is capable of delivering intricate responses to students swiftly, as demonstrated by the examples provided in \textbf{Table~\ref{tab:examples}}.

Our software is packaged as a Python library. LLM interactions are handled by the \textit{Guidance} Python package~\citep{microsoft_guidance}. User interfaces and a JSON HTTP API are supported by FastAPI \citep{fastapi}. We support traditional (e.g., JSON files, SQLite) as well as cloud-based data storage drivers. Our server can be run at low financial and computational cost through the combination of serverless deployment (e.g., to AWS Lambda) and serverless databases (e.g., AWS DynamoDB). Student responses are not stored by \textit{FreeText} infrastructure by default.

Any \textit{Guidance}-compatible LLM may be swapped into the Freetext server. That is, by default we access LLMs through the OpenAI API, but it is easy to swap in locally hosted or fine-tuned models: thus, privileged or sensitive information may be kept to on-premise compute resources, or users may opt to change which API-based LLM is accessed. For example, a more powerful LLM may be selected in cases where course content is particularly complex, or a simpler model may be used for more elementary course content. 

One front-end that students can access is a Jupyter Notebook widget, developed using IPyWidgets \citep{Kluyver2016jupyter}, making it easy to incorporate natural language short-answer questions as part of a notebook-based active-learning environment.

The widget communicates with the backend Python server described above. The widget is designed to be easily integrated into lecture and homework notebooks, enabling instructors to easily enrich existing teaching materials. A distinctive feature of our system is the intermediary server which equips the large language model with `held-out' information, such as a rubric for correct responses, accessible only to the LLM and instructor, and not to the student. This establishes the useful informational asymmetry between the evaluator and the student. 

To include the widget in a Python environment, the instructor can include the following code:

\begin{Verbatim}[samepage=true]
!pip install freetext_jupyter
from freetext_jupyter import FreetextWidget

FreetextWidget(
    # This ID is generated by the instructor.
    "07b2c3ef-0f97-46bc-a11e-..."
) 
\end{Verbatim}

When executed in a Jupyter notebook cell, this code will access the HTTP API to replace the widget with the corresponding question text for the student. Upon encountering the widget in a notebook, the student is presented with an open-ended question accompanied by a text box for response input. When they submit their response, the system transmits it to the server for combination with the feedback criteria set by the instructor.

\begin{figure*}[!ht]
    \includegraphics[width=\linewidth]{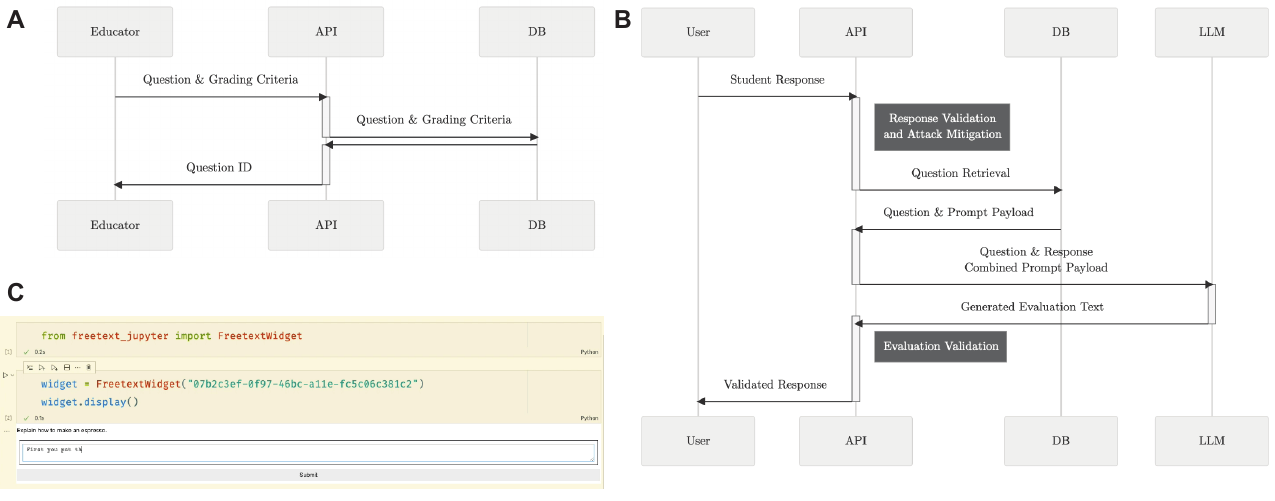}
    \caption{\textbf{A sequence diagram illustrating the flow of information within the \textit{FreeText} system.} \textbf{A.} First, an instructor formulates a question by supplying a student-facing question (``Question'') along with grading criteria for the LLM to evaluate student responses. In return, the educator obtains a unique identifier from the database, instrumental in retrieving the question text in the following step. \textbf{B.} Equipped with a unique Question identifier, a student provides an answer to the educator's query (``Response''). The API receives this request, pairing the Response with a Prompt based upon the educator's question and criteria, and directs them towards a large language model for evaluation. \textbf{C.} A screenshot of the \textit{FreeText} Jupyter widget integrated into an interactive code notebook.}
    \label{fig:sequence}
\end{figure*}

In the next stage, the student response and the pre-defined feedback criteria are bundled into a payload dispatched to a large language model. The LLM processes this payload and produces personalized feedback to the response. This feedback is relayed back to the student with seconds of latency through the web or notebook interface, offering them the immediate opportunity to reflect, amend, and improve their response as desired (\textbf{Fig.~\ref{fig:sequence}}).



Our tool is designed to be easily deployable and scalable. The \textit{FreeText} server can be run in resource-constrained or serverless platforms such as AWS Lambda. This allows for easy deployment and scaling, which is particularly important for large-scale projects and massive-scale courses~\citep{neuromatch}. Our API can also be combined with other existing educational tools in order to capture and store student responses for instructor review.

\subsection*{Question Design}

Instructors can provide a question for students to answer --- either programmatically, by accessing our HTTP API --- or graphically in the browser using the simple web application UI. Instructors can also provide optional assessment criteria --- text like \textit{``make sure the student mentions DNA base pairs in their answer.''}

\textit{FreeText} can use question content to automatically establish grading criteria, or it can use the assessment criteria to improve the text of the question. The latter process works by asking the AI to serve as a student and answer a question while oblivious to the instructor's grading criteria. Then, the answer is automatically evaluated by a separate instantiation of the LLM --- this time, against the instructor criteria. The assessment model determines if the student has been unfairly penalized due to omission of requirements (or a lack of clarity) in the original question text. If so, the question is updated to better encompass the requirements of the grading criteria. 

This process of iteratively incorporating assessment criteria is subtly different from simply including the criteria in the question text: For example, if the question text is, \textit{``What is the Rosetta Stone?''} and the criteria include, \textit{``Mention why the Ptolemaic dynasty created the Rosetta Stone''}, a \textit{bad} question update would be to explicitly ask about the Egyptian political system, as this gives the student more information than the instructor originally intended. A \textit{better} question update would be \textit{``Explain what the Rosetta Stone is and the context of its creation,''} because this nudges the student to discuss the right material but does not give any new information.

\subsection*{Question Presentation}

There are two built-in methods to present questions to students: the first is a simple web API, which can be used standalone, coupled with response-collection tools, or embedded within other web applications. The second is a Jupyter Notebook widget that can be embedded in tutorial coding notebooks.

The JSON web API endpoints may be accessed directly by application code, or students can access a simple web user interface. This interface comprises a question display and a textbox for student responses (see \textit{Supplemental Materials}). Feedback to students is rendered beneath the response box upon answer submission, and students may reuse the same page to resubmit amended answers.

The Jupyter Notebook widget is designed to make it easy for instructors to include open-ended questions in their assignments and subject the grading of student responses to custom grading criteria. This flexibility makes it easy for instructors to tailor the tool to their specific needs and teaching style.

\subsection*{Feedback to Students}

Our tool provides two types of feedback to students. The first is a holistic text response that provides feedback on the entire answer as a whole. The second is span-bound feedback (referring to a specific substring of the response) that can be used to highlight specific parts of the text that are erroneous or otherwise need student attention. For example, if a student's answer is correct but they misattribute a quote, the \textit{FreeText} server could highlight the attribution specifically to give feedback. The type of feedback returned can be specified by the instructor during question creation.




\section*{Discussion}

Here we introduced \textit{FreeText}, a framework capable of defining questions, collecting student responses, transmitting these responses alongside instructor expectations to a large language model (LLM), and generating rapid and personalized feedback for the students. Notably, the entirety of the student-facing workflow can be encapsulated within a Jupyter notebook, facilitating real-time enhancement of students' understanding of the course material. \textit{FreeText} is not confined to a web application and Jupyter notebooks, or the academic subjects mentioned above. The \textit{FreeText} Server can integrate with any application that consumes a JSON HTTP API, expanding its potential to a wider range of educational settings. 

Our system's broad applicability becomes evident when considering diverse learning models, such as the pod-based approach adopted by the online course Neuromatch Academy~\citep{neuromatch} in the field of computational neuroscience. In such settings, small student groups or `pods' collaboratively tackle assignments and projects. Teaching Assistants, tasked with providing feedback, can benefit from our tool, as it can streamline grading processes, reducing potential for attentional errors and freeing up instructors to deliver more personalized guidance to students.

Fully automated student evaluation is challenging both from a technical perspective and from a human perspective, and thus \textit{FreeText} is designed not to fully automate grading, but to serve as a useful tool benefiting both students and instructors. \textit{FreeText} benefits students by providing rapid and personalized feedback on short-answer questions. \textit{FreeText} benefits instructors by helping them to design better questions and grading criteria, by providing first-pass material for learning assessments, and by alleviating some of the burden of providing individualized instruction in large classes. LLMs in general, and \textit{FreeText} specifically, are not a replacement human instructors, but they can nonetheless fill a niche among education technologies.

LLMs undoubtedly hold immense power and potential. However, it is crucial to have an in-depth discussion about their ethical implications, especially in education. A key issue to consider is the potential biases that LLMs can introduce. These biases could unintentionally touch on sensitive subjects or unintentionally overlook marginalized groups. Instructors have a role to play by carefully designing their questions and assessment criteria. Further, students should be made aware of the nature of the system they are interacting with and its potential to make mistakes or act on internalized biases \citep{hsu2021attitudes}. On the other hand, automated systems such as \textit{FreeText} present an opportunity to reduce instructors' unconscious biases by evaluating all students' responses equally and without any explicit identification.

Furthermore, we must consider the broader dynamics of the AI ecosystem. The realm of LLMs is not limited to the offerings of large AI conglomerates like OpenAI. A burgeoning industry of alternative LLMs, both from smaller commercial entities and open-source initiatives \citep{claude2023,alpaca,llama2,huggingface}, is flourishing. Our framework is designed to be model-agnostic and can be 
readily adapted to integrate these alternative models.

Reliance solely on models from a single entity such as OpenAI raises two significant concerns. First, it centralizes the concentration of AI development resources and power, thereby exacerbating the already pronounced inequalities in the global AI landscape. Second, it can lead to a homogenization of the knowledge and perspectives propagated by AI models, potentially resulting in a limited and biased worldview. \textit{FreeText} is therefore \textit{deliberately} agnostic to the underlying LLM model and technologies.

We intend for our tool to enrich and expand students' educational experience, particularly in large-scale or resource-constrained course settings where detailed human intervention may be limited. Ongoing work includes the careful critique and evaluation of \textit{FreeText} outputs by expert instructors, taking advantage of upcoming opportunities to apply this technology in a large class setting.

Embracing both technical as well as human diversity helps mitigate many of the concerns raised above and enriches the AI ecosystem. A broad range of perspectives stalls the monopolization of AI technology and fosters a more balanced, equitable, and robust AI landscape. This viewpoint aligns with our belief in the need for broad and diverse human inputs, both in the creation of AI models and in their applications in society.

\begin{table*}[]
    \caption{Feedback examples.}
    \includegraphics[width=\linewidth]{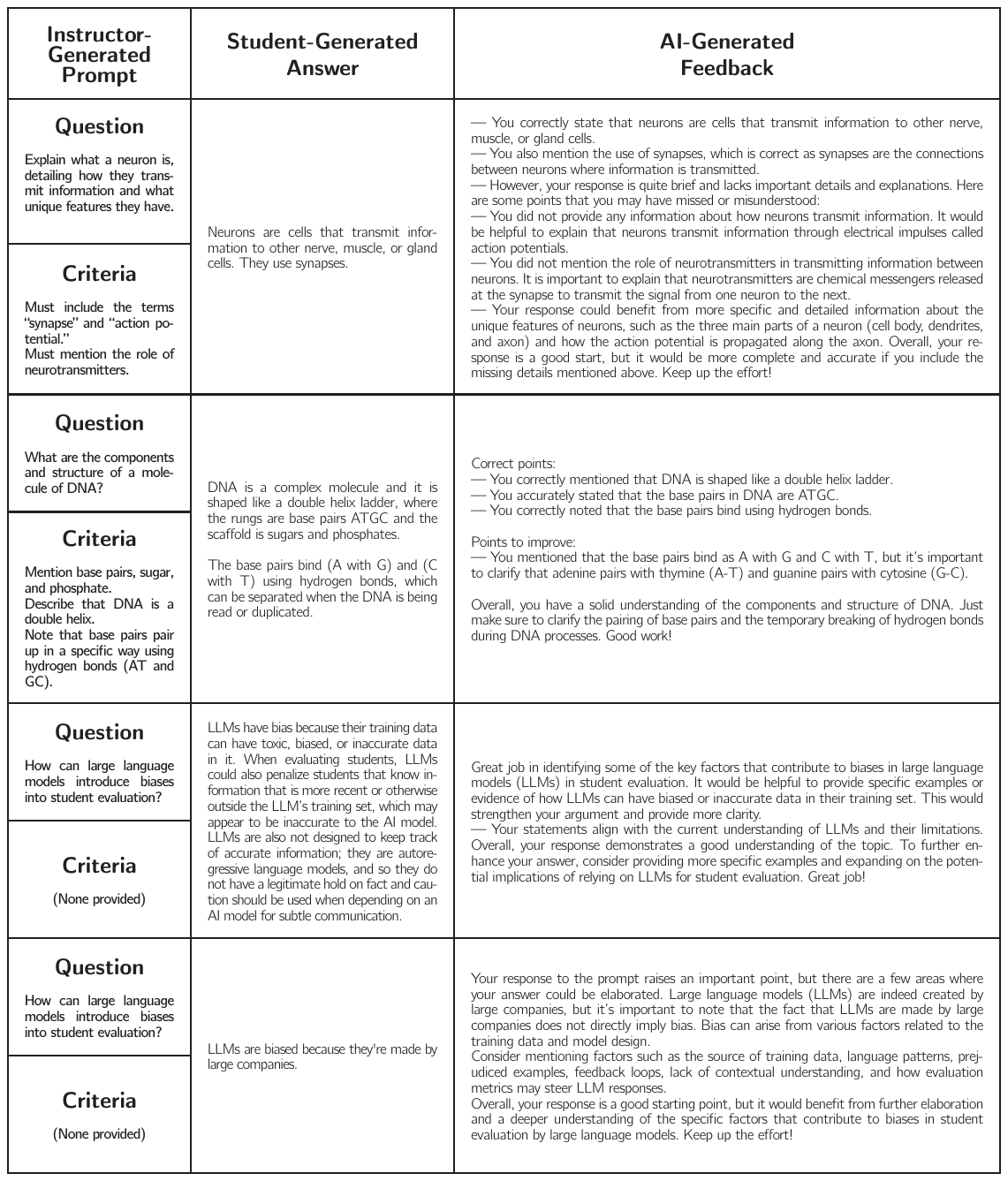}
    \label{tab:examples}
\end{table*}

\section*{Supplemental Materials}

Full-resolution versions of all images and tables from this publication are available at \url{https://llm4edu.experiments.kordinglab.com/paper}.

The FreeText server will be hosted temporarily for public use at \url{https://llm4edu.experiments.kordinglab.com/app}, with an interactive example assignment available at \url{https://llm4edu.experiments.kordinglab.com/app/assignments/1393754a-d80f-474d-bff7-b1fec36cdbb7}. Educators may contact us at the correspondence email of this preprint for a token, which is required to create new questions on our public instance.

Our Jupyter Notebook Widget is available on GitHub at \url{https://github.com/KordingLab/freetext-jupyter}, and is powered by the FreeText Server, which can be found at \url{https://github.com/KordingLab/llm4teach-freetext-server}.

\section*{Acknowledgements}

Research in this publication was supported by the National Institutes of Health under award number UC2-NS128361. The content is solely the responsibility of the authors and does not necessarily represent the official views of the National Institutes of Health.

\section*{Bibliography}
\bibliography{references.bib}

\begin{thebibliography}{30}
\providecommand{\natexlab}[1]{#1}
\providecommand{\url}[1]{\texttt{#1}}
\expandafter\ifx\csname urlstyle\endcsname\relax
  \providecommand{\doi}[1]{doi: #1}\else
  \providecommand{\doi}{doi: \begingroup \urlstyle{rm}\Url}\fi

\bibitem[Anthropic(2023)]{claude2023}
Anthropic.
\newblock {Claude}, 2023.
\newblock URL \url{https://www.anthropic.com}.
\newblock Accessed: 24 July 2023.

\bibitem[Brown et~al.(2020)Brown, Mann, Ryder, Subbiah, Kaplan, Dhariwal,
  Neelakantan, Shyam, Sastry, Askell, Agarwal, Herbert-Voss, Krueger, Henighan,
  Child, Ramesh, Ziegler, Wu, Winter, Hesse, Chen, Sigler, Litwin, Gray, Chess,
  Clark, Berner, {McCandlish}, Radford, Sutskever, and
  Amodei]{brown_language_2020}
T.~Brown, B.~Mann, N.~Ryder, M.~Subbiah, J.~D. Kaplan, P.~Dhariwal,
  A.~Neelakantan, P.~Shyam, G.~Sastry, A.~Askell, S.~Agarwal, A.~Herbert-Voss,
  G.~Krueger, T.~Henighan, R.~Child, A.~Ramesh, D.~Ziegler, J.~Wu, C.~Winter,
  C.~Hesse, M.~Chen, E.~Sigler, M.~Litwin, S.~Gray, B.~Chess, J.~Clark,
  C.~Berner, S.~{McCandlish}, A.~Radford, I.~Sutskever, and D.~Amodei.
\newblock Language models are few-shot learners.
\newblock In \emph{Advances in Neural Information Processing Systems},
  volume~33, pages 1877--1901. Curran Associates, Inc., 2020.

\bibitem[Burrows et~al.(2015)Burrows, Gurevych, and Stein]{burrows2015eras}
S.~Burrows, I.~Gurevych, and B.~Stein.
\newblock The eras and trends of automatic short answer grading.
\newblock \emph{International journal of artificial intelligence in education},
  25:\penalty0 60--117, 2015.

\bibitem[Cao(2023)]{cao_leveraging_2023}
C.~Cao.
\newblock Leveraging large language model and story-based gamification in
  intelligent tutoring system to scaffold introductory programming courses: A
  design-based research study, 2023.

\bibitem[Chung et~al.(2022)Chung, Hou, Longpre, Zoph, Tay, Fedus, Li, Wang,
  Dehghani, Brahma, Webson, Gu, Dai, Suzgun, Chen, Chowdhery, Castro-Ros,
  Pellat, Robinson, Valter, Narang, Mishra, Yu, Zhao, Huang, Dai, Yu, Petrov,
  Chi, Dean, Devlin, Roberts, Zhou, Le, and Wei]{chung_scaling_2022}
H.~W. Chung, L.~Hou, S.~Longpre, B.~Zoph, Y.~Tay, W.~Fedus, Y.~Li, X.~Wang,
  M.~Dehghani, S.~Brahma, A.~Webson, S.~S. Gu, Z.~Dai, M.~Suzgun, X.~Chen,
  A.~Chowdhery, A.~Castro-Ros, M.~Pellat, K.~Robinson, D.~Valter, S.~Narang,
  G.~Mishra, A.~Yu, V.~Zhao, Y.~Huang, A.~Dai, H.~Yu, S.~Petrov, E.~H. Chi,
  J.~Dean, J.~Devlin, A.~Roberts, D.~Zhou, Q.~V. Le, and J.~Wei.
\newblock Scaling instruction-finetuned language models, 2022.

\bibitem[Dzikovska et~al.(2013)Dzikovska, Nielsen, Brew, Leacock, Giampiccolo,
  Bentivogli, Clark, Dagan, and Dang]{dzikovska2013semeval}
M.~O. Dzikovska, R.~Nielsen, C.~Brew, C.~Leacock, D.~Giampiccolo,
  L.~Bentivogli, P.~Clark, I.~Dagan, and H.~T. Dang.
\newblock Semeval-2013 task 7: The joint student response analysis and 8th
  recognizing textual entailment challenge.
\newblock In \emph{Second Joint Conference on Lexical and Computational
  Semantics (* SEM), Volume 2: Proceedings of the Seventh International
  Workshop on Semantic Evaluation (SemEval 2013)}, pages 263--274, 2013.

\bibitem[Finnie-Ansley et~al.(2022)Finnie-Ansley, Denny, Becker, Luxton-Reilly,
  and Prather]{finnie2022robots}
J.~Finnie-Ansley, P.~Denny, B.~A. Becker, A.~Luxton-Reilly, and J.~Prather.
\newblock The robots are coming: Exploring the implications of openai codex on
  introductory programming.
\newblock In \emph{Proceedings of the 24th Australasian Computing Education
  Conference}, pages 10--19, 2022.

\bibitem[Heilman and Madnani(2013)]{heilman_ets_2013}
M.~Heilman and N.~Madnani.
\newblock {ETS}: Domain adaptation and stacking for short answer scoring.
\newblock In \emph{Second Joint Conference on Lexical and Computational
  Semantics (*{SEM}), Volume 2: Proceedings of the Seventh International
  Workshop on Semantic Evaluation ({SemEval} 2013)}, pages 275--279.
  Association for Computational Linguistics, 2013.

\bibitem[Hollingsworth(1960)]{hollingsworth_automatic_1960}
J.~Hollingsworth.
\newblock Automatic graders for programming classes.
\newblock \emph{Communications of the ACM}, 3\penalty0 (10):\penalty0 528--529,
  1960.
\newblock ISSN 0001-0782.
\newblock \doi{10.1145/367415.367422}.

\bibitem[Hsu et~al.(2021)Hsu, Li, Zhang, Fowler, Zilles, and
  Karahalios]{hsu2021attitudes}
S.~Hsu, T.~W. Li, Z.~Zhang, M.~Fowler, C.~Zilles, and K.~Karahalios.
\newblock Attitudes surrounding an imperfect ai autograder.
\newblock In \emph{Proceedings of the 2021 CHI conference on human factors in
  computing systems}, pages 1--15, 2021.

\bibitem[Kluyver et~al.(2016)Kluyver, Ragan-Kelley, P{\'e}rez, Granger,
  Bussonnier, Frederic, Kelley, Hamrick, Grout, Corlay, Ivanov, Avila, Abdalla,
  and Willing]{Kluyver2016jupyter}
T.~Kluyver, B.~Ragan-Kelley, F.~P{\'e}rez, B.~Granger, M.~Bussonnier,
  J.~Frederic, K.~Kelley, J.~Hamrick, J.~Grout, S.~Corlay, P.~Ivanov, D.~Avila,
  S.~Abdalla, and C.~Willing.
\newblock Jupyter notebooks -- a publishing format for reproducible
  computational workflows.
\newblock In F.~Loizides and B.~Schmidt, editors, \emph{Positioning and Power
  in Academic Publishing: Players, Agents and Agendas}, pages 87 -- 90. IOS
  Press, 2016.

\bibitem[Kortemeyer(2023)]{kortemeyer_can_2023}
G.~Kortemeyer.
\newblock Can an {AI}-tool grade assignments in an introductory physics
  course?, 2023.

\bibitem[Lathkar(2023)]{fastapi}
M.~Lathkar.
\newblock Getting started with fastapi.
\newblock In \emph{High-Performance Web Apps with FastAPI: The Asynchronous Web
  Framework Based on Modern Python}, pages 29--64. Springer, 2023.

\bibitem[Liu et~al.(2019)Liu, Wang, Wang, and Wu]{liu_automatic_2019}
X.~Liu, S.~Wang, P.~Wang, and D.~Wu.
\newblock Automatic grading of programming assignments: An approach based on
  formal semantics.
\newblock In \emph{2019 {IEEE}/{ACM} 41st International Conference on Software
  Engineering: Software Engineering Education and Training ({ICSE}-{SEET})},
  pages 126--137, 2019.
\newblock \doi{10.1109/ICSE-SEET.2019.00022}.

\bibitem[Messer et~al.(2023)Messer, Brown, Kölling, and
  Shi]{messer_automated_2023}
M.~Messer, N.~C.~C. Brown, M.~Kölling, and M.~Shi.
\newblock Automated grading and feedback tools for programming education: A
  systematic review, 2023.

\bibitem[Microsoft(2023)]{microsoft_guidance}
Microsoft.
\newblock Guidance.
\newblock \url{https://github.com/microsoft/guidance}, 2023.
\newblock Accessed: 24 July 2023.

\bibitem[Mizumoto and Eguchi(2023)]{mizumoto_exploring_2023}
A.~Mizumoto and M.~Eguchi.
\newblock Exploring the potential of using an {AI} language model for automated
  essay scoring.
\newblock \emph{Research Methods in Applied Linguistics}, 2\penalty0
  (2):\penalty0 100050, 2023.
\newblock ISSN 2772-7661.
\newblock \doi{10.1016/j.rmal.2023.100050}.

\bibitem[Mizumoto et~al.(2019)Mizumoto, Ouchi, Isobe, Reisert, Nagata, Sekine,
  and Inui]{mizumoto_analytic_2019}
T.~Mizumoto, H.~Ouchi, Y.~Isobe, P.~Reisert, R.~Nagata, S.~Sekine, and K.~Inui.
\newblock Analytic score prediction and justification identification in
  automated short answer scoring.
\newblock In \emph{Proceedings of the Fourteenth Workshop on Innovative Use of
  {NLP} for Building Educational Applications}, pages 316--325. Association for
  Computational Linguistics, 2019.
\newblock \doi{10.18653/v1/W19-4433}.

\bibitem[Orr and Russell(2021)]{orr_automatic_2021}
J.~W. Orr and N.~Russell.
\newblock Automatic assessment of the design quality of python programs with
  personalized feedback.
\newblock \emph{arXiv preprint arXiv:2106.01399}, 2021.

\bibitem[Radford et~al.(2019)Radford, Wu, Child, Luan, Amodei, and
  Sutskever]{radford_language_2019}
A.~Radford, J.~Wu, R.~Child, D.~Luan, D.~Amodei, and I.~Sutskever.
\newblock Language models are unsupervised multitask learners.
\newblock 2019.

\bibitem[Riordan et~al.(2017)Riordan, Horbach, Cahill, Zesch, and
  Lee]{riordan_investigating_2017}
B.~Riordan, A.~Horbach, A.~Cahill, T.~Zesch, and C.~M. Lee.
\newblock Investigating neural architectures for short answer scoring.
\newblock In \emph{Proceedings of the 12th Workshop on Innovative Use of {NLP}
  for Building Educational Applications}, pages 159--168. Association for
  Computational Linguistics, 2017.
\newblock \doi{10.18653/v1/W17-5017}.

\bibitem[Saha et~al.(2019)Saha, Dhamecha, Marvaniya, Foltz, Sindhgatta, and
  Sengupta]{saha2019joint}
S.~Saha, T.~I. Dhamecha, S.~Marvaniya, P.~Foltz, R.~Sindhgatta, and
  B.~Sengupta.
\newblock Joint multi-domain learning for automatic short answer grading.
\newblock \emph{arXiv preprint arXiv:1902.09183}, 2019.

\bibitem[Sung et~al.(2019)Sung, Dhamecha, Saha, Ma, Reddy, and
  Arora]{sung_pre-training_2019}
C.~Sung, T.~Dhamecha, S.~Saha, T.~Ma, V.~Reddy, and R.~Arora.
\newblock Pre-training {BERT} on domain resources for short answer grading.
\newblock In \emph{Proceedings of the 2019 Conference on Empirical Methods in
  Natural Language Processing and the 9th International Joint Conference on
  Natural Language Processing ({EMNLP}-{IJCNLP})}, pages 6071--6075.
  Association for Computational Linguistics, 2019.
\newblock \doi{10.18653/v1/D19-1628}.

\bibitem[Taori et~al.(2023)Taori, Gulrajani, Zhang, Dubois, Li, Guestrin,
  Liang, and Hashimoto]{alpaca}
R.~Taori, I.~Gulrajani, T.~Zhang, Y.~Dubois, X.~Li, C.~Guestrin, P.~Liang, and
  T.~B. Hashimoto.
\newblock Alpaca: A strong, replicable instruction-following model.
\newblock \emph{Stanford Center for Research on Foundation Models.
  https://crfm. stanford. edu/2023/03/13/alpaca. html}, 3\penalty0
  (6):\penalty0 7, 2023.

\bibitem[Touvron et~al.(2023)Touvron, Martin, Stone, Albert, Almahairi, Babaei,
  Bashlykov, Batra, Bhargava, Bhosale, et~al.]{llama2}
H.~Touvron, L.~Martin, K.~Stone, P.~Albert, A.~Almahairi, Y.~Babaei,
  N.~Bashlykov, S.~Batra, P.~Bhargava, S.~Bhosale, et~al.
\newblock Llama 2: Open foundation and fine-tuned chat models.
\newblock \emph{arXiv preprint arXiv:2307.09288}, 2023.

\bibitem[Ureel~{II} and Wallace(2019)]{ureel_ii_automated_2019}
L.~C. Ureel~{II} and C.~Wallace.
\newblock Automated critique of early programming antipatterns.
\newblock In \emph{Proceedings of the 50th {ACM} Technical Symposium on
  Computer Science Education}, {SIGCSE} '19, pages 738--744. Association for
  Computing Machinery, 2019.
\newblock ISBN 978-1-4503-5890-3.
\newblock \doi{10.1145/3287324.3287463}.

\bibitem[van Viegen et~al.(2021)van Viegen, Akrami, Bonnen, DeWitt, Hyafil,
  Ledmyr, Lindsay, Mineault, Murray, Pitkow, et~al.]{neuromatch}
T.~van Viegen, A.~Akrami, K.~Bonnen, E.~DeWitt, A.~Hyafil, H.~Ledmyr, G.~W.
  Lindsay, P.~Mineault, J.~D. Murray, X.~Pitkow, et~al.
\newblock Neuromatch academy: Teaching computational neuroscience with global
  accessibility.
\newblock \emph{Trends in cognitive sciences}, 25\penalty0 (7):\penalty0
  535--538, 2021.

\bibitem[Wolf et~al.(2020)Wolf, Debut, Sanh, Chaumond, Delangue, Moi, Cistac,
  Ma, Jernite, Plu, Xu, Le~Scao, Gugger, Drame, Lhoest, and Rush]{huggingface}
T.~Wolf, L.~Debut, V.~Sanh, J.~Chaumond, C.~Delangue, A.~Moi, P.~Cistac, C.~Ma,
  Y.~Jernite, J.~Plu, C.~Xu, T.~Le~Scao, S.~Gugger, M.~Drame, Q.~Lhoest, and
  A.~M. Rush.
\newblock {Transformers: State-of-the-Art Natural Language Processing}.
\newblock pages 38--45. Association for Computational Linguistics, Oct. 2020.
\newblock URL \url{https://www.aclweb.org/anthology/2020.emnlp-demos.6}.

\bibitem[Yoon(2023)]{yoon_short_2023}
S.-Y. Yoon.
\newblock Short answer grading using one-shot prompting and text similarity
  scoring model, 2023.

\bibitem[Zhu et~al.(2022)Zhu, Wu, and Zhang]{zhu2022automatic}
X.~Zhu, H.~Wu, and L.~Zhang.
\newblock Automatic short-answer grading via bert-based deep neural networks.
\newblock \emph{IEEE Transactions on Learning Technologies}, 15\penalty0
  (3):\penalty0 364--375, 2022.

\end{thebibliography}

\end{document}